\documentclass[twocolumn,prl,superscriptaddress]{revtex4-1}
\usepackage{amsmath,amssymb,mathrsfs}
\usepackage{natbib}
\usepackage{subfigure}
\usepackage{tabularx}
\usepackage{epsfig}
\usepackage{longtable}
\usepackage{amsfonts}
\usepackage{rotating}
\usepackage{bbold}
\usepackage{hhline}
\usepackage{braket}
\usepackage{txfonts, comment}
\usepackage{multirow}
\usepackage{appendix}
\usepackage[unicode=true,bookmarks=true,bookmarksnumbered=false,bookmarksopen=false,breaklinks=false,pdfborder={0 0 1},backref=false,colorlinks=true]{hyperref}

\hypersetup{linkcolor=magenta,urlcolor=blue,citecolor=blue,pdfstartview={FitH},hyperfootnotes=false,unicode=true}

\def\be{\begin{equation}}
\def\ee{\end{equation}}
\def\bea{\begin{eqnarray}}
\def\eea{\end{eqnarray}}

\begin{document}

\title{Diffusive transport from spatially correlated random phase kicks}


\author{Pei Wang}
\affiliation{Department of Physics, Zhejiang Normal University, Jinhua 321004, China}
\email{wangpei@zjnu.cn}

\begin{abstract}
We study the dynamics of a single-particle wave packet on a one-dimensional lattice subject to periodic random phase kicks with finite spatial correlation length. This stroboscopic setting provides a controllable model of dephasing in driven quantum systems. Using a momentum-space formulation, we show that the evolution is governed by an accumulated phase whose structure determines the spreading of the wave packet. We find that the phase kicks strongly suppress ballistic transport and induce diffusion at long times. We derive an explicit analytical expression for the diffusion coefficient as a function of the correlation length, in excellent agreement with numerical simulations. Our results uncover a simple mechanism by which spatially correlated phase noise controls quantum transport, and provide a quantitatively testable prediction for diffusion in periodically driven lattice systems. Possible experimental realizations in cold-atom platforms are discussed.
\end{abstract}


\maketitle

\section{Introduction}
\label{sec:intro}

The transport of wave packets in the presence of temporal 
noise has been extensively studied over the past several decades~\cite{Madhukar77,Jayannavar82,Ott84,Evensky90,Lebedev95,Krivolapov12,Gopalakrishnan17,Rossi17,Rath20,Bhakuni24,Longhi24}. 
For a particle on a one-dimensional lattice, where the quasi-momentum 
is bounded by the lattice structure, it is well understood that 
noise acting on onsite potentials induces dephasing. Such dephasing 
suppresses coherent interference and converts ballistic transport 
into diffusive behavior~\cite{Madhukar77,Evensky90,Gopalakrishnan17}. 
Starting from a localized initial state, the wave-packet 
width exhibits quadratic growth in time in the absence 
of noise, while it crosses over to linear growth under dephasing.

Most previous studies have focused on continuous-time dynamics. 
In this framework, the evolution is governed either by a Hamiltonian 
with explicitly time-dependent stochastic terms~\cite{Evensky90,Lebedev95,Krivolapov12}, or equivalently by a 
Lindblad equation in which the noise is incorporated as a 
dephasing channel acting on the density matrix~\cite{Rath20,Bhakuni24,Longhi24}. These approaches 
are well suited for describing electron transport in solid-state 
systems, where particles are continuously coupled to environmental 
degrees of freedom such as phonons.

However, such noise-driven transport is difficult to control in a 
quantitative and tunable manner. In particular, while theoretical 
predictions for the diffusion coefficient $D$ have been established 
in various models, direct experimental verification remains 
challenging due to the lack of precise control over noise strength and correlation properties.

An alternative approach is to consider stroboscopic dynamics with 
periodic phase kicks. Such protocols can be realized in a highly 
controllable manner, for example in cold-atom systems. When the 
kicks are random, they effectively mimic temporal noise, with the 
accumulated phase during each period corresponding to an onsite 
potential fluctuation. Despite this close connection, the dynamics 
of wave packets under random phase kicks has received relatively 
limited attention~\cite{Ott84}.

In this work, we introduce a minimal lattice model in which random 
phase kicks are applied periodically. Importantly, the phase 
kicks are spatially correlated: rather than being independent at each site, 
they are characterized by a finite correlation length 
$\xi$. Specifically, the phase difference between two sites $x$ and 
$x'$ follows a Gaussian distribution with the variance
$(x - x')^2 / \xi^2$. This construction defines a spatially correlated 
noise field and allows us to systematically explore the 
role of correlation length in transport.

We develop a momentum-space framework to analyze the resulting 
dynamics. Within this approach, the wave function can be expressed 
analytically in momentum space, and the effect of phase kicks is 
captured through the accumulation of a $k$-dependent phase. 
This enables us to directly relate real-space spreading to the statistical 
properties of the phase gradient. Using this method, 
we show that the wave packet exhibits diffusive spreading at long times 
and derive an explicit expression for the diffusion coefficient,
$D \propto \frac{1}{e^{1/(2\xi^2)} - 1}$,
which differs qualitatively from previously reported results.

Finally, we discuss possible experimental realizations of our model. 
The random phase kicks considered here are equivalent to 
applying a time-dependent random force to particles on a lattice. 
Periodic driving with controlled forces is already well 
established in several platforms, including cold atoms in optical lattices~\cite{Dahan96,Geiger18}, 
synthetic dimensions~\cite{Oliver23}, photonic waveguide arrays~\cite{Morandotti99}, and coupled nanocavities~\cite{Kimura10}. 
In particular, constant forces leading to Bloch oscillations have been 
widely demonstrated. Extending these setups to include 
randomized driving provides a feasible route to test our predictions, 
including the dependence of the diffusion coefficient 
on the noise correlation length.

The remainder of this paper is organized as follows. 
In Sec.~\ref{sec:model}, we introduce the phase-kick model. 
In Sec.~\ref{sec:nosu}, we present numerical results for the wave-packet dynamics. 
In Sec.~\ref{sec:momen}, we develop a momentum-space formulation and use it to analyze the dynamics. 
In Sec.~\ref{sec:diff}, we show that random phase kicks give rise to diffusive transport and derive 
an explicit expression for the diffusion coefficient. 
Finally, in Sec.~\ref{sec:con}, we discuss possible experimental realizations.

\section{Model}
\label{sec:model}

We consider a one-dimensional lattice described by the Hamiltonian
$\hat{H}_0 = - g \sum_{x} \left( \hat{c}^\dag_x \hat{c}_{x+1} + \mathrm{h.c.}\right)$,
where $g$ is the hopping amplitude and $\hat{c}_x$ annihilates a particle at site $x$.
The dynamics is defined stroboscopically. Over one period $T$, the wave function evolves as
\begin{equation}\label{eq:M:ev}
\ket{\psi_{t+1}} = \exp \left\{-i \sum_x \theta_x \hat{c}^\dag_x \hat{c}_x \right\}
\exp \left\{-i T \hat{H}_0 \right\} \ket{\psi_t}.
\end{equation}
Each period consists of a coherent evolution generated by $\hat{H}_0$, followed by a phase kick.

The phase $\theta_x$ is taken to be a spatially correlated Gaussian field,
\begin{equation}\label{eq:M:th}
\theta_x = \frac{1}{\sqrt{N}\,\xi} \sum_{x'} (x - x')\, W_{x'},
\end{equation}
where $N$ is the total number of lattice sites (assumed large), and $W_{x'}$ 
are independent standard normal random variables. This construction ensures 
translational invariance and that the statistical properties of $\theta_x$ are independent of $N$.
In particular, the phase difference between two sites satisfies
\begin{equation}
\overline{(\theta_x - \theta_{x'})^2} = \frac{(x - x')^2}{\xi^2}.
\end{equation}
For separations $\left|x - x' \right| \ll \xi$, the phase difference is negligible, indicating 
strong phase coherence. In contrast, for $\left|x - x' \right| \gg \xi$, the variance becomes 
large and the relative phase is effectively random. Thus, phase coherence is 
preserved only within a characteristic length scale $\xi$, which we refer to as the correlation length.

We study the evolution of an initial Gaussian wave packet,
\begin{equation}
\psi_0(x) \propto \exp\left[
-\frac{(x - x_0)^2}{4\sigma_0^2} + i k_0 x
\right],
\end{equation}
where $\sigma_0$ and $k_0$ denote the initial width and momentum, respectively. 
The state is evolved by iterating Eq.~\eqref{eq:M:ev}, and we analyze the spreading 
of the wave packet as a function of the discrete time $t$. The phase kicks at different 
time steps are taken to be statistically independent.
This setup allows us to directly probe the interplay between coherent lattice dynamics 
and spatially correlated phase fluctuations.

For completeness, we briefly comment on the density-matrix description. 
Defining the noise-averaged density matrix as
$\hat{\rho}_t = \overline{\ket{\psi_t}\bra{\psi_t}}$,
one finds that the phase kicks induce a dephasing factor in real space,
$\rho_t(x,y) \rightarrow e^{-(x-y)^2/(2\xi^2)} \, \rho_t(x,y)$,
which suppresses off-diagonal coherence over distances larger than $\xi$. 
In this work, however, we focus on the wave-function dynamics, which contains 
more detailed information about the underlying transport process.

\section{Noise-induced suppression of wave spreading}
\label{sec:nosu}

\begin{figure}[htp]
\centering
\vspace{0.1cm}
\includegraphics[width=.48\textwidth]{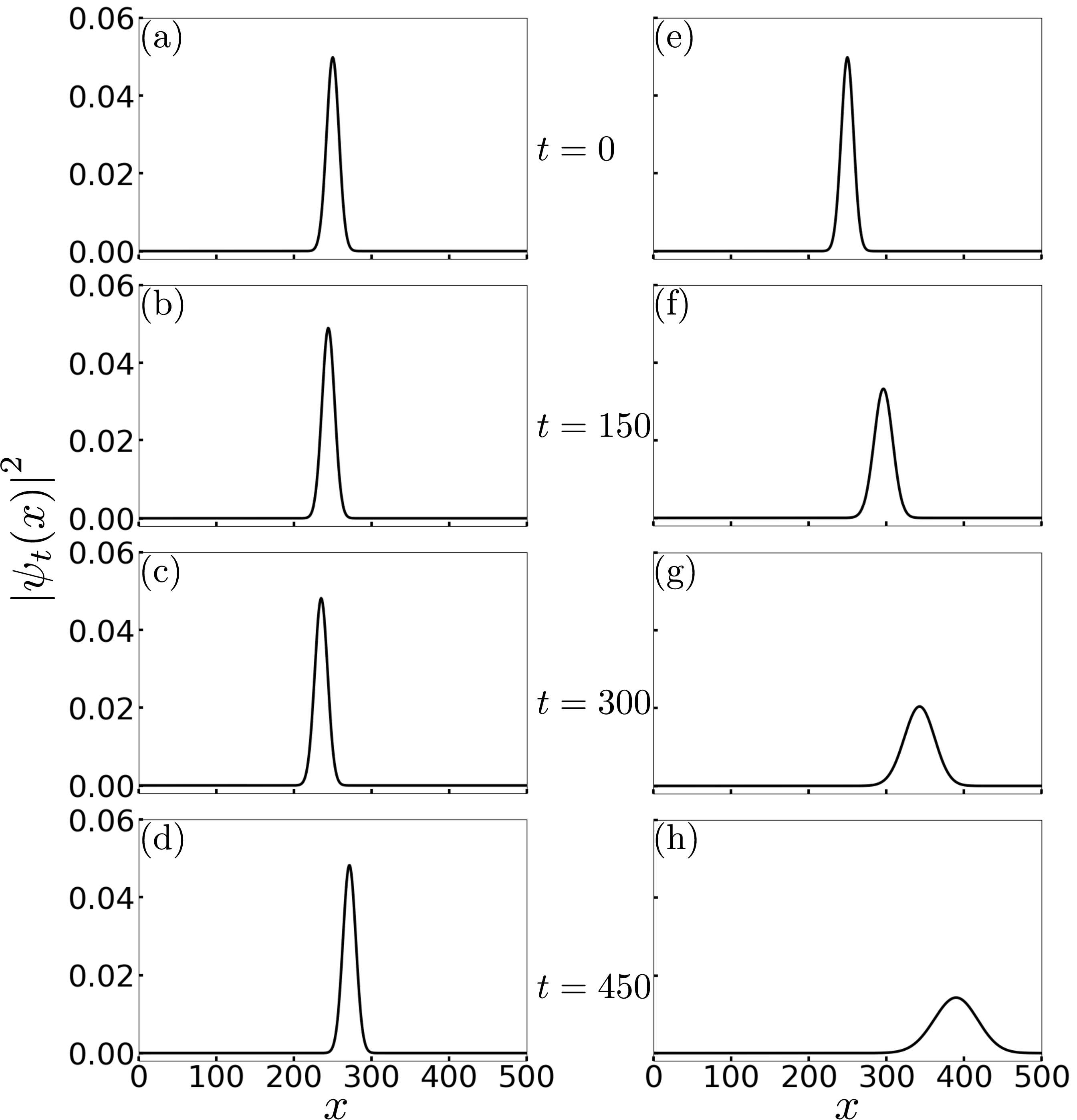}
\caption{Snapshots of the probability density $\left|\psi_t(x)\right|^2$ at times 
$t = 0, 150, 300, 450$. Panels (a)--(d): evolution under random phase 
kicks with correlation length $\xi = 1$. Panels (e)--(h): free evolution without phase kicks. 
In both cases, the initial state is a Gaussian wave packet with 
$\sigma_0 = 8$ and $k_0 = \pi/10$. We use $N = 500$ lattice sites with 
periodic boundary conditions to avoid boundary effects. 
The phase kicks strongly suppresses the spreading and destroys directed motion.}
\label{fig:snapshot}
\end{figure}

We first examine the real-space dynamics of the wave packet. 
Figure~\ref{fig:snapshot} shows snapshots of the probability density 
$|\psi_t(x)|^2$ at representative times. Under pure free evolution,
$\ket{\psi_{t+1}} = e^{-i T \hat{H}_0} \ket{\psi_t}$,
the wave packet retains an approximately Gaussian profile while its width 
increases with time (right panels of Fig.~\ref{fig:snapshot}). Meanwhile, 
the center of the packet propagates with a constant velocity determined 
by the initial momentum $k_0$. In Fig.~\ref{fig:snapshot}, we choose 
$k_0 = \pi/10$, such that the packet moves steadily to the right.

In contrast, when phase kicks are introduced according to 
Eq.~\eqref{eq:M:ev}, the spreading of the wave packet is strongly suppressed 
(left panels of Fig.~\ref{fig:snapshot}). Over the same time interval, the 
width shows only a weak increase compared to the free evolution case. At 
the same time, the motion of the wave-packet center becomes irregular: the 
initial momentum $k_0$ is rapidly lost, and the center undergoes a random 
motion with no preferred direction. On long time scales, the packet remains 
localized near its initial position, without exhibiting systematic drift.

This behavior indicates a strong suppression of transport induced by the 
phase kicks. As discussed in the previous section, phase coherence is 
preserved only within a distance $\xi$, while correlations are lost at 
larger scales. As a result, long-range interference is suppressed, leading 
to inhibited spreading.

We emphasize that the qualitative features described above are robust and 
do not depend sensitively on the specific values of $g$, $T$, $\sigma_0$, 
or $k_0$. In the simulations shown here, we set $2gT = 1.0$. 
In contrast, the dynamics depends strongly on the correlation length $\xi$, 
which controls the strength of phase decoherence.

\section{Momentum space mechanism}
\label{sec:momen}

The suppression of wave-packet spreading can be understood 
by transforming to momentum space. We define the momentum-space wave function as
$\psi_t(k) = \frac{1}{\sqrt{N}} \sum_x \psi_t(x) e^{-ikx}$,
where $k \in [-\pi, \pi)$ is the quasi-momentum.
For the initial Gaussian wave packet, one obtains (within a continuum approximation)
$\psi_0(k) \propto \exp\left\{ -\sigma_0^2 (k - k_0)^2 - i x_0 k \right\}$,
up to an overall normalization factor. The structure of the exponent directly encodes 
the properties of the wave packet: the real quadratic term determines the width, 
while the imaginary linear term determines the center position.

This representation is particularly convenient because the time evolution can be 
expressed in a simple form. The phase kick defined in Eq.~\eqref{eq:M:th} can be 
rewritten, up to an irrelevant constant, as $\theta_x = q x$, where
\begin{equation}
q = \frac{1}{\sqrt{N}\,\xi} \sum_{x'} W_{x'}
\end{equation}
is a Gaussian random variable with zero mean and variance $1/\xi^2$. 
Each phase kick thus induces a momentum shift $k \to k + q$.
Since the free Hamiltonian $\hat{H}_0$ is diagonal in momentum space, the combined 
evolution over one period takes the form $\psi_{t+1}(k) = e^{i 2gT \cos(k+q)} \, \psi_t(k+q)$.
Iterating this relation, the wave function at time $t$ can be written as
\begin{equation}
\psi_t(k) \propto \exp\left\{ -\sigma_0^2 (k - \tilde{k}_t)^2 - i \tilde{F}_t(k) \right\}.
\end{equation}

The effective momentum $\tilde{k}_t = k_0 - \sum_{\tau=1}^t q_{\tau}$
characterizes the instantaneous momentum of the wave packet and thus its velocity. 
Since $q_{\tau}$ are independent Gaussian variables with zero mean and variance $1/\xi^2$, 
$\tilde{k}_t$ performs a random walk in momentum space with 
$\mathrm{Var}(\tilde{k}_t) = t/\xi^2$. Because quasi-momentum is defined modulo $2\pi$, 
$\tilde{k}_t$ is effectively confined to $[-\pi,\pi)$. At long times, repeated random 
shifts lead to a uniform distribution over this interval. As a result, the momentum---and 
hence the velocity---becomes completely randomized, explaining the absence of directed 
motion observed numerically.

The accumulated phase is
\begin{equation}
\tilde{F}_t(k) = x_0 k - 2gT \sum_{j=1}^t \cos\!\left(k + \sum_{\tau=j}^t q_{\tau} \right).
\end{equation}
The real-space properties of the wave packet are governed by the $k$-dependence of 
this phase. In particular, the derivative
\begin{equation}
X(k) = \frac{d \tilde{F}_t}{dk}
\end{equation}
determines the spatial structure: its average gives the center position, while its 
variation over $k$ controls the width.

\begin{figure}[htp]
\centering
\vspace{0.1cm}
\includegraphics[width=.48\textwidth]{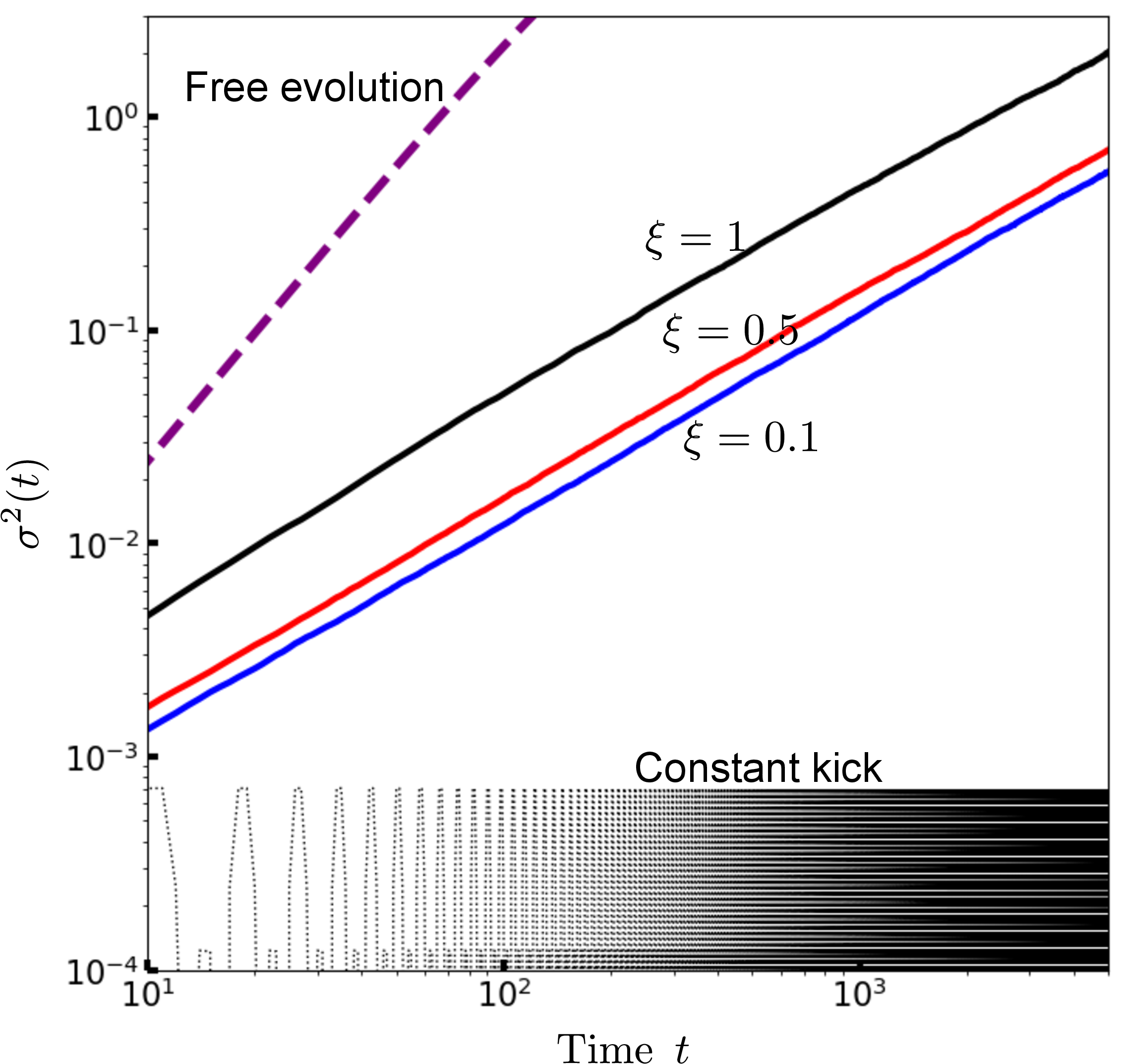}
\caption{Squared width of the wave packet, $\sigma^2(t)$, as a function of time. 
The purple dashed line corresponds to ballistic spreading under free evolution, 
while the black dotted line shows dynamical localization for a constant momentum kick. 
The solid curves represent random phase kicks with correlation lengths 
$\xi = 1.0$ (black), $\xi = 0.5$ (red), and $\xi = 0.1$ (blue), illustrating diffusive behavior.}
\label{fig:sigma2}
\end{figure}

In the present model, $X(k)$ takes the form
\begin{equation}
X(k) = x_0 + \tilde{S}_I \cos k + \tilde{S}_R \sin k,
\end{equation}
where $\tilde{S}_R$ and $\tilde{S}_I$ quantify the variation of $X(k)$ over momentum. 
Since $X(k)$ is a bounded trigonometric function, its range is set by the magnitude 
of these coefficients. Consequently, the wave-packet width obeys
\begin{equation}\label{eq:diff:s2}
\sigma^2(t) \;\propto\; \tilde{S}_R^2(t) + \tilde{S}_I^2(t),
\end{equation}
up to some constant prefactor. Here, the squared width is defined as
$\sigma^2(t) = \overline{\bra{\psi_t} \left(\hat{x} - x(t) \right)^2 \ket{\psi_t}}$,
where $x(t) = \bra{\psi_t} \hat{x} \ket{\psi_t}$ is the center position, and 
the overline denotes averaging over noise realizations. The scaling behavior of 
$\sigma^2(t)$ is therefore directly governed by the time dependence of 
$\tilde{S}_R^2 + \tilde{S}_I^2$.

The role of these coefficients can be illustrated in two limiting cases.

\emph{(i) Free evolution.}  
In the absence of phase kicks ($q_{\tau} \equiv 0$), one finds
$\tilde{S}_I = 0$ and $\tilde{S}_R = 2gT\, t$.
Thus, the variation of $X(k)$ grows linearly in time, leading to 
$\sigma^2(t) \propto t^2$, characteristic of ballistic transport. 
This behavior is confirmed in Fig.~\ref{fig:sigma2} (purple dashed line).

\emph{(ii) Constant momentum kick.}  
If $q_{\tau} \equiv q$ is a constant, $\tilde{S}_R$ and $\tilde{S}_I$ become periodic 
functions of time with period $2\pi/q$. As a result, $X(k)$ remains bounded, and 
the wave-packet width exhibits oscillatory behavior without long-time growth. 
This corresponds to dynamical localization~\cite{Dunlap86,Holthaus95}. In Fig.~\ref{fig:sigma2}, the black 
dotted line shows periodic oscillations of $\sigma^2(t)$, confirming the absence 
of spreading.

These examples demonstrate that wave-packet spreading is controlled by the growth 
of the phase gradient in momentum space. In the next section, we show that random 
phase kicks lead to a diffusive growth of $\tilde{S}_R^2 + \tilde{S}_I^2$, and 
hence to diffusive transport.


\section{Diffusive transport and diffusive coefficient}
\label{sec:diff}

\begin{figure}[htp]
\centering
\vspace{0.1cm}
\includegraphics[width=0.43\textwidth]{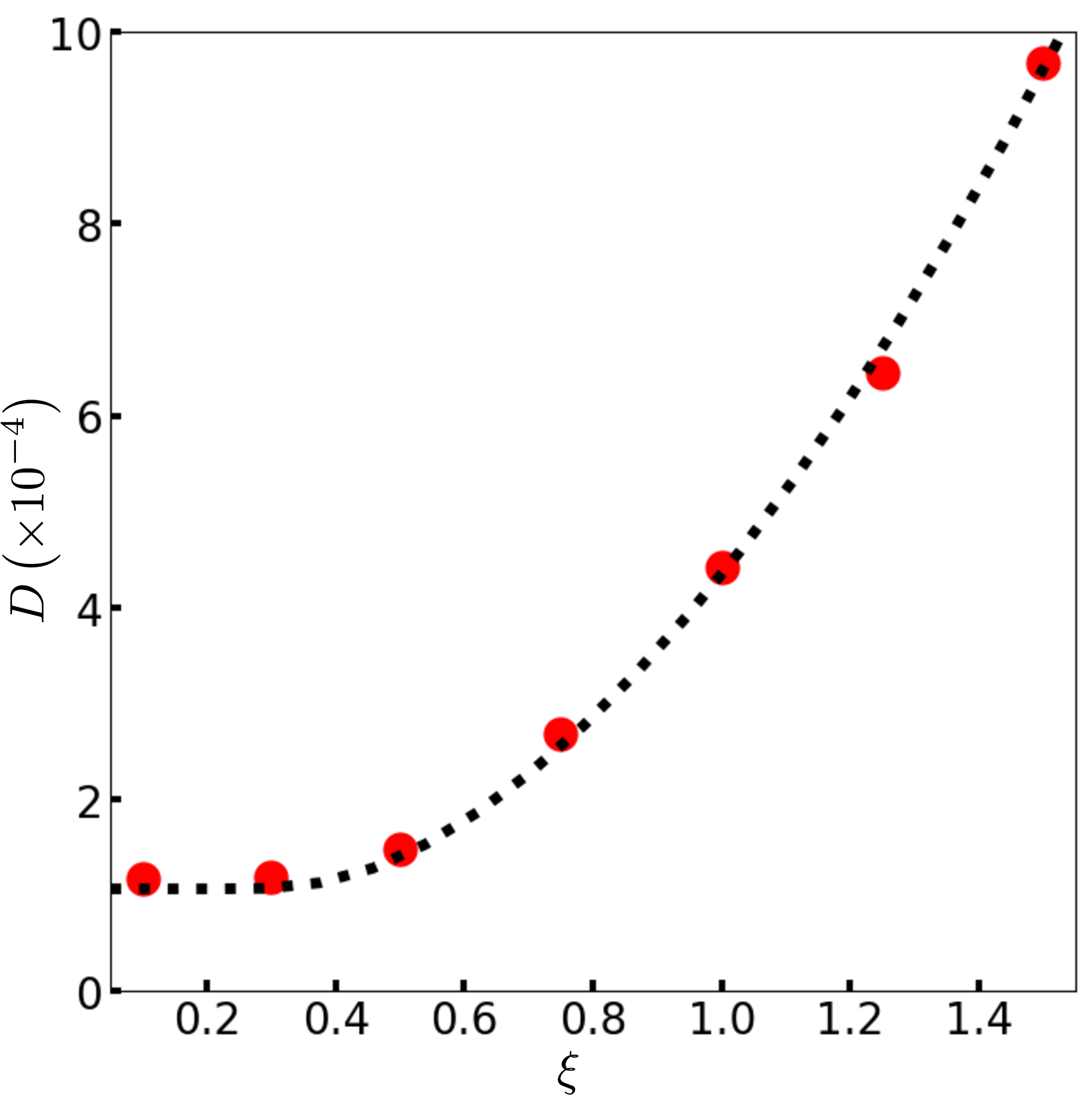}
\caption{Diffusion coefficient $D$ as a function of the correlation length $\xi$. 
The black dotted line shows the analytical prediction of Eq.~\eqref{eq:diff:D}, 
while the red symbols represent numerical results extracted from the long-time 
growth of $\sigma^2(t)$. Excellent agreement is observed between theory and numerics.}
\label{fig:D}
\end{figure}

Under random phase kicks, the variables $q_{\tau}$ ($\tau = 1,2,\dots,t$) are 
independent and identically distributed Gaussian random variables. As a result, 
the coefficients $\tilde{S}_R$ and $\tilde{S}_I$ become random variables. It is 
convenient to combine them into a complex quantity 
$\tilde{S} = \tilde{S}_R + i \tilde{S}_I = 2gT\, S$, where
\begin{equation}\label{eq:app:S}
S = e^{iq_t} + e^{i(q_t + q_{t-1})} + \cdots + e^{i(q_t + q_{t-1} + \cdots + q_1)}.
\end{equation}

To evaluate the statistical properties of $S$, we use the identity valid for a 
Gaussian random variable $q$, $\overline{e^{iq}} = e^{-\mathrm{Var}(q)/2}$,
where the overline denotes averaging over noise realizations. This relation 
allows us to compute the moments of $S$.

In the long-time limit, the mean of $S$ approaches a constant,
\begin{equation}\label{eq:app:ES}
\lim_{t\to\infty} \overline{S} = \frac{1}{e^{1/(2\xi^2)} - 1},
\end{equation}
which already provides a good approximation when $t/\xi^2 \gg 1$. In contrast, 
the second moment grows with time. A straightforward calculation yields, for 
$t/\xi^2 \gg 1$,
\begin{equation}
\overline{|S|^2} = t + \frac{2(t-1)}{e^{1/(2\xi^2)} - 1}
- \frac{2}{\left(e^{1/(2\xi^2)} - 1\right)^2}.
\end{equation}
Thus, $\overline{|S|^2}$ increases linearly with $t$ at long times.

Since $\tilde{S} = 2gT\, S$, it follows that
$\overline{\tilde{S}_R^2 + \tilde{S}_I^2} = (2gT)^2 \, \overline{|S|^2}$,
and therefore also grows linearly in time. Using the relation established in the 
previous section, $\sigma^2(t) \propto \tilde{S}_R^2 + \tilde{S}_I^2$, we conclude 
that the squared width exhibits diffusive scaling, $\sigma^2(t) \propto t$,
which is the hallmark of diffusive transport. This linear growth originates from 
the accumulation of random phase gradients, which perform an effective random 
walk in momentum space.

We now verify this prediction numerically. Figure~\ref{fig:sigma2} shows 
$\sigma^2(t)$ as a function of $t$, obtained by averaging over $10^4$ independent 
noise realizations. For all values of the correlation length $\xi$ (solid black, 
red, and blue curves), $\sigma^2(t)$ exhibits a clear linear dependence on time, 
in agreement with the analytical result. The data are presented on logarithmic 
scales to clearly distinguish different scaling behaviors, including ballistic 
($\sigma^2 \propto t^2$), diffusive ($\sigma^2 \propto t$), and bounded dynamics 
associated with dynamical localization.

To quantify the transport, we introduce the diffusion coefficient $D$, defined by
$\sigma^2(t) = 2Dt$. From the above analysis, we obtain
\begin{equation}\label{eq:diff:D}
D \propto \frac{1}{2} + \frac{1}{e^{1/(2\xi^2)} - 1},
\end{equation}
up to a $\xi$-independent prefactor.

Equation~\eqref{eq:diff:D} is the central result of this work. It provides an 
explicit dependence of the diffusion coefficient on the correlation length, 
which differs qualitatively from conventional dephasing-induced diffusion.

The diffusion coefficient $D$ is a monotonically increasing function of $\xi$. 
In the limit $\xi \to 0$, $D$ approaches a finite constant. This limit corresponds 
to spatially uncorrelated phase kicks, where each lattice site experiences an 
independent random phase. Such spatially white noise leads to standard diffusive 
transport, consistent with previous studies of dephasing-induced diffusion.

In contrast, in the large-$\xi$ limit, Eq.~\eqref{eq:diff:D} predicts 
$D \propto 2\xi^2$, indicating a rapid increase of the diffusion coefficient. 
Physically, increasing $\xi$ enhances the spatial correlation of the phase kicks, 
so that sites within a region of size $\xi$ experience nearly identical phase shifts. 
As a result, phase coherence is partially preserved over this length scale, 
reducing the effective dephasing and allowing more efficient transport. 
In the extreme limit $\xi \to \infty$, the phase becomes spatially uniform, 
corresponding to a coherent momentum shift rather than true dephasing, and the 
diffusive description is expected to break down in favor of coherent dynamics.

In Fig.~\ref{fig:D}, we compare the analytical prediction~\eqref{eq:diff:D} 
(black dotted line) with numerical results for $D$ (red symbols), extracted 
from the long-time slope of $\sigma^2(t)$. The agreement is excellent over the 
entire range of $\xi$, confirming the validity of our theory. The overall 
prefactor is obtained by fitting the numerical data to Eq.~\eqref{eq:diff:D}.


\section{Discussion: experimental realization}
\label{sec:con}

Finally, we discuss a possible experimental realization of the phase-kick model. 
The phase kick takes the form $\theta_x = qx$, which is linear in the spatial 
coordinate and therefore equivalent to a linear potential. Its effect can be 
interpreted as a spatially uniform force $F$ acting on the particle.
In our stroboscopic formulation~\eqref{eq:M:ev}, $x$ is dimensionless and takes 
integer values. For a physical lattice with lattice constant $a$, the phase 
accumulated during a finite time interval $\tau$ under a constant force $F$ is 
related to $q$ via $qx = -\frac{F a x \tau}{\hbar}$.
Thus, a phase kick corresponds to briefly switching on a constant force.

The required dynamics can be implemented by periodically alternating between 
free evolution and force application. Specifically, each period of duration $T$ 
consists of two steps: (i) free evolution under $\hat{H}_0$ for time $T-\tau$, 
and (ii) evolution under a constant force $F$ for a short time $\tau$. To realize 
the stochastic phase kicks, the force amplitude is chosen independently in each 
period according to a Gaussian distribution.

A natural platform for implementing this protocol is provided by cold atoms in 
optical lattices, where constant forces and Bloch oscillations have been widely 
studied~\cite{Dahan96,Geiger18}. In such systems, the lattice potential is generated by a standing wave 
of laser light, and a controllable force can be introduced by imposing a 
time-dependent frequency difference $\delta\nu(t)$ between the two 
counter-propagating beams. A time-dependent ramp $\delta\nu(t)$ generates a force
$F = -\frac{m\lambda}{2}\,\dot{\delta\nu}(t)$,
where $\lambda$ is the laser wavelength and $m$ is the atomic mass~\cite{Dahan96}. The lattice 
constant is $a = \lambda/2$, leading to the relation
$q = \frac{m\lambda^2 \tau}{4\hbar}\,\dot{\delta\nu}(t)$.

In our model, $q$ is drawn from a Gaussian distribution with variance $1/\xi^2$. 
This can be implemented by programming the ramp rate $\dot{\delta\nu}(t)$ to be 
a random variable with the corresponding variance. In this way, the correlation 
length $\xi$ can be tuned experimentally through the statistical properties of 
the applied force, enabling direct measurement of the predicted dependence of 
the diffusion coefficient on $\xi$.

Beyond cold atoms, similar protocols may be implemented in photonic waveguide 
arrays or coupled resonator systems, where effective lattice dynamics and 
Bloch oscillations have also been realized. We leave a detailed exploration of 
these platforms for future work.

\end{document}